\documentclass{ws-ijmpd}

\usepackage[super,compress]{cite}
\usepackage{enumerate}
\usepackage{soul}
\usepackage[normalem]{ulem}
\usepackage{xcolor}

\begin{document}

\markboth{P. Bosso, S. Das}
{On Lorentz invariant mass and length scales}

\catchline{}{}{}{}{}
%
%%%%%%%%%%%%%%%%%%%%%%%%%%%%%%%%%%%%%%%%%%%%%%%%%%%%%%%%%%%%%%%%%%%%

\title{ON LORENTZ INVARIANT MASS AND LENGTH SCALES}

\author{PASQUALE BOSSO
\footnote{Address from 01 September 2018: %\newline
Divisi\'on de Ciencias e Ingenier\'ias,%\newline 
Universidad de Guanajuato,%\newline
Loma del Bosque 103, Lomas del Campestre CP 37150, Le\'on, Gto., Mexico. %\newline
Email:bosso@fisica.ugto.mx }
%\footnote{Typeset names in
%8~pt roman, uppercase. Use the footnote to indicate the
%present or permanent address of the author.}
}

\address{
Theoretical Physics Group and Quantum Alberta, \\
Department of Physics and Astronomy, University of Lethbridge, \\  
4401 University Drive, Lethbridge, Alberta, Canada T1K 3M4
\\
pasquale.bosso@uleth.ca}

\author{SAURYA DAS}

\address{
Theoretical Physics Group and Quantum Alberta, \\
Department of Physics and Astronomy, University of Lethbridge, \\  
4401 University Drive, Lethbridge, Alberta, Canada T1K 3M4
\\
saurya.das@uleth.ca}

\newcommand{\tcr}[1]{\textcolor{red}{#1}}
\newcommand{\tcb}[1]{\textcolor{blue}{#1}}

\maketitle

\begin{history}
%\received{Day Month Year}
%\revised{Day Month Year}
\end{history}

\begin{abstract}
We show that the standard Lorentz transformations admit an invariant mass (length) scale, such as the Planck scale.
In other words, the frame independence of such scale is built-in within those transformations, and one does not need to invoke the principle of relativity for their invariance. 
This automatically ensures the frame-independence of 
the spectrum of geometrical operators in quantum gravity. 
Furthermore, we show that the above predicts a small but measurable
difference between the inertial and gravitational mass of any object, 
regardless of its size or whether it is elementary or composite. 

\end{abstract}


\ccode{PACS numbers: 04.60.-m, 04.60.Bc, 03.30.+p}

%\tableofcontents

It is generally argued that quantum gravitational effects will be 
manifest when the energy of a system is close to Planck energy, i.e. 
\mbox{$E \simeq M_\mathrm{P} c^2 = \sqrt{\hbar c^5/G}~(\approx 10^{16}$ TeV}),
equivalently when probed distances are close to Planck length scale, 
i.e. \mbox{$L \simeq \ell_\mathrm{P} = \sqrt{G \hbar/c^3}~(\approx 10^{-35}$ m}).
For example, string degrees of freedom are expected to show up near that scale, as
are minimum measurable lengths, areas and volumes\cite{string} . 
Similar results hold for discrete areas and volumes in loop quantum gravity\cite{lqg} . 
Note that $E$ and $L$ are the 
{\it physical} energy and length, which are {\it frame-dependent}
and transform under Lorentz transformations (e.g. due to usual Lorentz contraction).
On the other hand, $M_\mathrm{P}$ and $\ell_\mathrm{P}$ are simply scales, which together with their constituents $c$, $\hbar$, and $G$, are \emph{Lorentz invariant}.
%%%
The Lorentz transformations follow from the 
{\it Principle of Relativity} (POR), 
which postulates the equivalence of all 
vacua, or inertial frames.
%%%%
Then the invariance of $c$ follows directly from the Lorentz transformations themselves. 
However that of $\hbar$ and $G$ are argued 
directly from the POR, 
%{\it principle of relativity} (equivalence of inertial frames), 
since any variation in them can be used to look
for a preferred inertial frame, effectively bringing back the concept of aether. 

%\tcr{However, the standard dispersion relation, coupled with the standard Lorentz transformations, results in the standard composition law.
%A theory so structured has the usual problem of physical length and mass scales (i.e. $E$ and $L$ respectively) not being Lorentz invariant.}
In an attempt to make these physical scales and quantum gravity effects frame-independent, such that no preferred inertial frame is singled out, the so-called ``Doubly Special Relativity" (DSR) theories were first proposed in\cite{ac1,ac1b} , explored in\cite{glikman} 
and developed in\cite{dsr1,dsr2} .
In this, a modified Lorentz transformation is postulated, which preserves 
both $c$ and $\ell_\mathrm{P}$.
However, this gives rise to new problems, such as the wavelength dependence of the velocity of light, {modified composition laws for momenta\cite{Judes2003}}~, its inapplicability to macroscopic systems (the so-called ``soccer ball problem''\cite{ac,hossenf,unruh}~, a possible solution of which is given in Ref. 12), and the non-standard energy-momentum dispersion relation\cite{ac}
\begin{eqnarray}
E^2= (\vec pc)^2 + (mc^2)^4 + f(E, \vec{p}^2, M_P).
\label{dsrdisp}
\end{eqnarray}
Eq.\eqref{dsrdisp} implies $dE/dp \equiv v (\vec p, m, M_P)$, energy $E \neq m c^2 \gamma$ and momentum $\vec p \neq m\vec v \gamma$, where $\gamma = (1-\beta^2)^{-1/2}$ and $\beta = v/c$, for a particle of mass $m$, moving at velocity $\vec v$.
This means that if two such non-interacting particles are simply held together (e.g. by a string), then $E_\mathrm{total} \neq E_1 + E_2$ and $\vec p_\mathrm{total} \neq \vec p_1 + \vec p_2 $.
That is, for certain laws of physics such
as dispersion relations, there is no single formula which holds for small and large masses, elementary and composite systems, which is clearly undesirable. 
Thus it is important to examine this issue further.
{It is worth mentioning that similar problems may arise for two and multi-particle 
systems even for certain theories 
with standard dispersion relations
\cite{Kowalski-Glikman2003,mignemi,carmona} .}

In this paper, we ask a slightly different but related question:
{\it Can one retain the standard Lorentz transformations,
as well as the standard energy-momentum dispersion relation, such that 
invariant mass and length scales are built-in
within the Lorentz transformations, just as $c$ is?}
This would mean that:
\begin{enumerate}[(i)]
\item $M_{\mathrm P}$, and by extension $\hbar$ and $G$, would be invariant by virtue of the Lorentz transformations alone, and
\item while physical energies, such as that of a high energy beam in vacuum,
would still be frame-dependent, this would guarantee that 
eigenvalues of quantum operators 
such as lengths and areas of a Lorentz-covariant theory
will be frame-independent. 
This may even partially or wholly resolve the issue of frame-dependence of quantum gravity effects \cite{speziale} .
Furthermore, as we will show later, relations such as Eq.\eqref{dsrdisp} above appear as {\it effective} low energy equations in our model; they will have the similar phenomenological implications, but without related problems referred to above (see e.g. Eq.\eqref{disp3} later in the paper).
\end{enumerate}

We show that the answer to the above question is surprisingly yes:
the standard Lorentz transformations indeed admit of an invariant mass scale (which may or may not be the Planck scale). 
Furthermore, we show that this result makes concrete observable predictions, in terms of 
a small but acceptable (i.e. as long as strict bounds are met \cite{will})
violation of the Equivalence Principle. 
Also in this case, there is no associated ``soccer-ball' type problem.
Theories with minimum length\cite{scardigli} as well as many modifications of general relativity\cite{teleparallel} also predict such violations. 

%We start by noting that in quantum gravity, there are two 
%independent and relevant dimensionless parameters,
%$\beta=v/c$ and $m/M_\mathrm{P}$.
%As shown in Figure 1, there are four distinct regimes \\
%
%(i) $\beta \ll 1, %\rightarrow 0, 
%m/M_\mathrm{P} \ll 1 %\rightarrow 0
%$ 
%(non-relativistic, classical gravitational)\\
%
%(ii) $\beta \rightarrow 1, 
%m/M_\mathrm{P} \ll 1 %\rightarrow 0
%$ 
%(relativistic, classical gravitational)\\
%
%(iii) $\beta \ll 1, %\rightarrow 0, 
%m/M_\mathrm{P} \gtrsim 1 %\rightarrow\infty 
%$ (non-relativistic, quantum gravitational)\\
%%
%(iv) $\beta \rightarrow 1, m/M_\mathrm{P} \gtrsim 1 %\rightarrow \infty
%$ (relativistic, quantum gravitational)

%In the above, `non-relativistic' refers to low-velocities of the source, which could nevertheless be massive and give rise to a strong gravitational field. 
%Further, in characterizations (iii) and (iv) above, we assume that the
%particle is elementary, although as we shall see, the subsequent 
%formalism holds good for composite objects as well. 

%\begin{figure}%[htp] 
%\centering
%{
%\includegraphics[width=0.4\textwidth]
%%{2scales.pdf}
%{Regimes5.png                                                                                                                                                              }
%}
%\caption{Depicting two independent tunable scales, $\beta$ and $m/M_P$}
%\end{figure}

%%\vspace{-1.5cm}
%%\noindent
%%Fig1: Depicting two independent tunable scales, $\beta$ and $m/M_P$

\vspace{0.5cm}
%Next, we examine if a {\it mass scale} $M$ can be built-in within the 
We start with the standard Lorentz transformations (along $x,x'$ axis):
\begin{eqnarray}
x'=\gamma(x-vt),~~t'=\gamma (t-vx/c^2)~.
\label{lt1}
\end{eqnarray}
The standard expression for energy and momentum and the $4$-momentum $p^a$, incorporating an arbitrary mass scale $M$ are:
\begin{align}
E = & \mu(m,M_\mathrm{}) c^2 \gamma ~, & \vec p = & \mu(m,M_\mathrm{}) \vec v~\gamma~,\label{em1} \\
p^a = & \left(\frac{E}{c},\vec p \right)~, \label{em2}
\end{align}
where 
\begin{eqnarray}
\mu = \mu(m,M_\mathrm{}) \label{mu1}
\end{eqnarray}
is a function of the parameter $m$ and 
mass scale $M$. We will show below that
$\mu=m + \mbox{\it higher order terms}$. 
%and the parameter $m$ corresponding to $\mu$ in %the limit of usual special relativity.
The standard dispersion relation holds 
\begin{eqnarray}
p_a p^a = \frac{E^2}{c^2} - p^2 = (\mu c)^2~, \label{disp1}
\end{eqnarray}
which together with Eq.\eqref{em1} implies that energy and momentum simply add up for non-interacting particles. 
As usual, $p_\alpha p^\alpha = (\mu c)^2$ is the Casimir of the Lorentz symmetry group. 
This implies that $\mu$, as well as $m$, 
which is the leading term in any expansion 
of $\mu$ in series of powers of $m/M$ or $M/m$ 
(see Eqs.(\ref{lim1}),(\ref{lim2}),(\ref{lim2a}) and following text for more details) are Lorentz invariants. 
Therefore it follows 
that the mass scale $M$ is Lorentz invariant as well. 
%
%
%Moreover, it incorporates the mass scale $M$, and therefore also the 
In other words, $M$ and the 
length scales $\ell_1 = \hbar/Mc$ and $\ell_2 = GM/c^2$ (and time scales $t_{1,2} = \ell_{1,2}/c$) 
are now incorporated in the Lorentz transformations and therefore automatically Lorentz invariant, just as desired. One does not need to appeal to the POR 
again for their invariance. 
%
%in a {\it Lorentz invariant} way, since 
%Eq.\eqref{disp1} is Lorentz invariant.
%In fact, $m$ has to be an invariant, since by definition it is the leading term of any expansion of $\mu$ in series of powers of $m/M$ or $M/m$ (see below for more details.)
%Our proposal, therefore, allows for the existence of an 
%
%Note that at these are {\it not necessarily} the Planck scales.
One can identify $M$ with the Planck mass,
although strictly speaking not necessary. 

We realize that $\mu$ is nothing but the inertial mass of the system, and that the replacement $m \rightarrow \mu$ would simply amount to re-naming the mass, and of no observational consequence.
{Experimentally, the difference between $\mu$ and $m$ may arise in physical processes which depends on the mass parameter $m$.
Equivalently, in a set of equations which depend on {\it both} $\mu$ and $m$.}
The simplest example is of course if one assumes
$m$ to be the {\it gravitational mass}
of the system, such that (for Newton's gravity with potential $\phi$)
\begin{eqnarray}
\frac{d(\mu \vec v)}{dt} =  - m \vec\nabla \phi~.
\end{eqnarray}
This would mean that the gravitational masses of two particles do not simply add up{, as the inertial masses do,} when they form a composite object.
This is inconsequential in the absence of gravitational interactions, for example in standard model processes, as there is no way to measure these masses.
If gravity is taken into account on the other hand, general relativity tells us that gravitational mass of the composite is not an algebraic sum of its parts.
Its predictions (computed numerically, for example) can then be used to determine the shape of the function $\mu$ in Eq. \eqref{mu1}, or equivalently the scales $m_0$ and coefficients $d_j$ in Eq. \eqref{lim} below. 

Therefore, as in the above, if $m$ is interpreted as the gravitational mass, 
the most stringent bounds for the difference between gravitational and inertial mass 
go as \cite{will,gundlach}
\begin{eqnarray}
\frac{|\mu-m|}{\mu} = 10^{-N_1},~~N_1 {\geq} 13 \label{ep1}~.
\end{eqnarray}
%
%\tcr
Note that in \cite{will}, the author considers violations of the equivalence principle resulting from various internal constituents of an object contributing differently to its gravitational mass.
In our case instead, the origin of such a violation is our proposed relation between inertial and gravitational masses for all objects
(Eq. \eqref{mu1} and Eqs. \eqref{lim1}, \eqref{lim2} below). 
However, the tests reviewed in \cite{will} would be sensitive to the overall difference between the two masses (if any), regardless of the origin of such a difference. 
The experiments giving rise to the above bound involve masses in a wide range, 
from the microscopic ($m \simeq 10^{-33} $ kg, $m/M_P \simeq 10^{-28}$)
%($m/M_P \simeq 10^{-28}$) 
to the macroscopic ($m \simeq 1$ kg, $m/M_P \simeq 10^8$).
%($m/M_P \simeq 10^8$). 
We therefore require the following behavior for $\mu(m,M)$
for all possible range of masses:
\begin{subequations} \label{lim}
\begin{align}
\mu = & m + m_0 \cdot {\cal O} \left(\frac{m}{M} \right)
~, & \mbox{as}~\frac{m}{M} \ll & 1 %\rightarrow 0 
\label{lim1} \\
\mu = & m + m_0 \cdot {\cal O} \left(\frac{M}{m} \right)
~, & \mbox{as}~\frac{m}{M} \gg & 1 % \rightarrow \infty~
, \label{lim2} \\
& \frac{m_0}{M} \ll 1~. \label{lim2a}
\end{align}
\end{subequations}
The above ensures that, consistent with experiments, the inertial and gravitational masses can differ at most by a tiny amount throughout the range of $\mu$ and $m$, and also that one can smoothly interpolate for intermediate mass ranges (i.e. $m/M={\cal O}(1)$), using Eqs. \eqref{lim1} and \eqref{lim2}.
These in turn guarantee that 
Eq.(\ref{mu1}) holds for all ranges of mass, small or big. 
As stated previously, from the Lorentz invariance of $\mu$ (Eq.(\ref{disp1})), the
form of Eqs.(\ref{lim1}) and (\ref{lim2}) above, and the linear and homogeneous nature of the Lorentz transformations (\ref{lt1}), it follows that both $m$ and $M$ are Lorentz invariant as well. 
%
%Now since both $\mu$ and the gravitational mass $m$ 
%are Lorentz invariant 
%(as otherwise weak gravitational interactions 
%in inertial frames can be used to pick out a special frame), 
%it follows that $m_0$ {\it as well as} the mass scale $M$ are
%Lorentz invariant, as promised earlier (the appearance of $m/M$ and its inverse $M/m$ in Eqs.(\ref{lim1}) and (\ref{lim2}) 
%respectively guarantees this). 
Note that the above requires the existence of a new mass scale, $m_0$, which should have observational consequences.
As mentioned earlier, 
if one constructs a quantum theory, e.g. including the gravitational field, 
then any Lorentz invariant spectrum of length, area etc. should be proportional to 
the above length scales, its squared etc. 
For example, if $M=M_P$, then one would get 
$\text{Length} = n \ell_P$, 
$\text{Area} = n \ell_P^2$, with $n \in \mathbb{N}$, with 
their minima $\ell_P$ and $\ell_P^2$ respectively \cite{lqg,speziale,adv0}.
Other attempts to construct a Lorentz invariant mass/length scale include 
\cite{dvaether,subir}.

Next as a concrete example, we consider the following function:
\begin{equation}
	\mu (m,M_\mathrm{}) = 
    m + m_0 \left[ \tanh \left( \frac{m}{M} 
\right)^a \right] \sum_{j=1}^\infty d_j \left( \frac{M_\mathrm{}}{m} \right)^j~,
\label{tanh1}
\end{equation}
where $a>1 \in \mathbb{R},~d_j =0, \forall j>a$~.
This indeed gives the following expansions:
\begin{subequations}
\begin{align}
\mu = & m + m_0 \left( \frac{m}{M_\mathrm{}} \right)^a 
\left[ 1 - \frac{1}{3} \left( \frac{m}{M} \right)^2 + \cdots \right] {\sum_{j=1}^{\infty} d_j \left( \frac{M_\mathrm{}}{m} \right)^j}, & 
\mbox{as} \quad & {\frac{m}{M} \ll 1}
\label{lim3}\\
\mu = & m + m_0 \left[  1 + 2 \sum_{l=0}^\infty (-)^l e^{- \frac{2lm}{~M} }
\right] 
\sum_{j=1}^{\infty } d_j \left( \frac{M}{m} \right)^j~, &
\mbox{as} \quad  & \frac{m}{M} \gg 1~. %\rightarrow \infty~.
\label{lim4} 
\end{align}
\end{subequations}
As can be seen, Eqs.(\ref{lim3}-\ref{lim4}) 
above are of the form of Eqs.(\ref{lim1}-\ref{lim2}) respectively. 

We now examine ways of testing the current proposal. 
In the range $m/M \ll 1$, neutron interferometry may provide the required clue.
For example, the relative phase shift between two interferometer paths at different elevations, as a result of a tilt of the interferometer by an angle $\alpha$ (subjecting the paths to slightly different gravitational fields)
is given by 
\begin{eqnarray}
\phi_g = \frac{\mu m g \lambda A\sin\alpha}{2\pi \hbar^2}~,
\label{cow1}
\end{eqnarray}
where $\mu$ ($m$) is the inertial (gravitational) mass of the neutrons, $\lambda$ their de Broglie wavelength, $g$ acceleration due to gravity and $A$ the area enclosed by the two paths \cite{werner,zeilinger}
(one can also use its relativistic counterpart \cite{ryder}).
The expansion for $\mu$ can be read off from Eq.(\ref{lim3}). 
If the accuracy of measurement in these set of experiments (or observed difference between theory and experiment) $\sim 10^{-N_2}$, then it follows from Eq.(\ref{lim3}) with $a=1$: 
\begin{eqnarray}
\frac{|\mu-m|}{m} = \frac{m_0}{M} \leq 10^{-N_2}
\label{bound1}
\end{eqnarray}
{Note that experiments such as the above were not originally designed for testing the equivalence principle, but can be used for such especially since the neutron mass in Eq.(\ref{cow1}) can be measured very accurately \cite{kruger}. 
It is in fact worth exploring if existing data have hidden within them any violations of the principle. 
}

In the range $m/M \gg 1$, there may be several ways of testing the proposal as well. 
We note that Eqs.(\ref{lim3}-\ref{lim4}) are already  
of the form of Eq.(1) of \cite{will} 
%\tcr
{($\mu = m + \sum_A \eta^A E^A/c^2$)}, 
which parametrizes
the violation of the equivalence principle, with the identifications 
$E^A/c^2 \simeq m_0$,
$\eta^A \simeq (m/M_P)^a$ and $\eta^A \simeq (M_P/m)^{j_{max}}$ respectively.  
However, here we present a theoretical basis of the origin of that equation, and show that it is valid for all objects, irrespective of composition, and for 
all range of masses. 
Therefore from \cite{will} and Eq.(\ref{lim4}) (for $a \in \mathbb{N}$), one gets
(for gravitational mass $m'$):  
\begin{eqnarray}
\frac{|\mu-m'|}{m'} = \frac{m_0 M}{m'^2}  \leq 10^{-N_1} ~.
\label{bound2}
\end{eqnarray}
%
%
%For example, if $m = m_e = 0.5 $ MeV, the electron mass, one gets from Eqs.(\ref{ep1})
%and (\ref{lim4}) (for $a \in \mathbb{N}$), 
%$m_0 \leq 10^{-34}~m_e$, or equivalently.
%
Bounds such as (\ref{bound1}) and (\ref{bound2}) from different experiments,
together with Eqs.(\ref{lim3})-(\ref{lim4}) can be 
combined to estimate both $m_0$ and $M$ as:
\begin{align}
m_0 & \leq 10^{-(N_1+N_2)/2}~m' \label{bound3} \\
10^{-(N_1+3N_2)/2}~m' &\leq M \leq 10^{(N_2-N_1)/2}~m'~, \label{bound4}
\end{align}
where to get Eq.(\ref{bound4}), we have used Eq.(\ref{bound1}) and 
Eq.(\ref{bound3}) as an equality.
%Note that gravitational masses, e.g. the 
%neutron mass $m$ in Eq.(\ref{cow1}) can be measured very accurately \cite{kruger}.
Note the following: 
\begin{enumerate}[(i)]
\item to test our proposal, the two test bodies used to define the E\"otv\"os ratios \cite{will} must have different gravitational masses, 
\item a non-null result for these tests with two objects of the same gravitational mass (but different inertial masses) will falsify our proposal.
\end{enumerate}

Once the existence of an invariant mass or length scale is established, 
one can employ effective theories to implement this minimum length. 
The Generalized Uncertainty Principle (GUP) and the associated deformed Heisenberg algebra is one such effective model of arriving at a minimum length \cite{kmm,adv,Bosso} . 
Although the algebra of operators therein is not manifestly Lorentz covariant, being merely an effective low energy model, this does not pose a problem.
Note that manifestly Lorentz covariant GUP algebras have been proposed \cite{quesne,ghosh} , and that the possibility of the underlying theory violating Lorentz invariance has been explored \cite{liberati} .
Here too, any potential issues associated with composite objects, e.g.
as detailed in \cite{ac2} is absent in reality, for the above reason.
Then, following the definition of \cite{adv} , we have 
(for one spatial dimension for simplicity):
\begin{eqnarray}
p = p_0 \left( 1 - \alpha p_0 + 2\alpha^2 p_0^2 \right),~
\alpha= \frac{\alpha_0}{M_P c}, 
~\alpha_0={\cal O}(1)~,
\end{eqnarray}
where $p_0$ is the effective ``low energy momentum'', with classical Poisson bracket
$\{x_i,p_{0j}\}=\delta_{ij}$ and quantum commutator bracket 
$[x_i, p_{0j}]=i\hbar \delta_{ij}$. 
Substituting this, along with the function $\mu(m)$ from \eqref{lim3} and \eqref{lim4} 
for $m/M_P \ll 1$
and $m/M_P \gg 1$ respectively, 
we get from Eq. \eqref{disp1}
\begin{subequations} \label{disp3}
\begin{align}
E^2 = & (p_0 c)^2 + (m c^2)^2 - 2\alpha p_0^3 + 5 \alpha^2 p_0^4 \nonumber \\
& + 2 m m_0 c^4 \left( \frac{m}{M_P} 
\right)^a
\left[ 1 - \frac{1}{3} \left( \frac{m}{M_P} \right)^2 + \cdots \right] {\sum_{j=1}^{\infty} d_j \left( \frac{M_P\mathrm{}}{m} \right)^j} \nonumber \\
	& + m_0^2 c^4 \left( \frac{m}{M_P} 
\right)^{2a}
\left[ 1 - \frac{1}{3} \left( \frac{m}{M_P} \right)^2 + \cdots \right]^2 \left[{\sum_{j=1}^{\infty} d_j \left( \frac{M_P\mathrm{}}{m} \right)^j}\right]^2,
\label{disp3a}
\\
E^2 =& (p_0 c)^2 + (m c^2)^2 - 2\alpha p_0^3 + 5 \alpha^2 p_0^4 \nonumber \\
& + 2 m m_0 c^4 
\left[  1 + 2 \sum_{l=0}^\infty (-)^l e^{- \frac{2lm}{~M_P} }
\right] 
\sum_{j=1}^{\infty } d_j \left( \frac{M_P}{m} \right)^j\nonumber \\
& + m_0^2 c^4 
\left[  1 + 2 \sum_{l=0}^\infty (-)^l e^{- \frac{2lm}{~M_P} }
\right]^2
\left[\sum_{j=1}^{\infty } d_j \left( \frac{M_P}{m} \right)^j\right]^2.
\label{disp3b}
\end{align}
\end{subequations}
Note that the RHS of Eq. \eqref{disp1} is in terms of $p$ and $\mu$, while that of Eqs. \eqref{disp3} are in terms of $p_0$ and $m$.
Therefore for example for astrophysical observations, with gravitational mass $m~(\gg M_P)$ being the relevant parameter, and noting that for quantum phenomena $\vec p_0 = -i\hbar \vec\nabla$, Eq.(\ref{disp3b}) above is practically of the same form as Eq.(\ref{dsrdisp}), with $\vec p \rightarrow \vec p_0$ and with some additional terms. 
Consequently, the phenomenological implications of these
two equations would also be similar, which may be used to estimate $m_0$ and $d_j$. 

Another set of potential astrophysical observations include
corrections to the perihelion precession of planets such as
Mercury, using the modified geodesic equations incorporating small violations of the equivalence principle
\cite{aldrovandi}:
\begin{eqnarray}
\frac{du^\mu}{d\tau} + \Gamma^\mu_{\nu\lambda} u^\nu u^\lambda 
= \left( \frac{m-\mu}{m} \right) \partial^\mu x_a \frac{du^a}{d\tau}~.
\end{eqnarray}
Detailed studies of these are left to a future publication. 

Finally, this formalism can easily be generalized to include
$n$ mass scales $M_1, M_2, \dots, M_P, \dots, M_n $, arranged in 
ascending order, with $2 n$ corresponding 
length scales $\ell_{k1} = \hbar/M_k c$,
$\ell_{k2}=GM_k/c^2$ ($k=1,\dots,n$).
% (and $2n$ time scales $t_{i1,2}=\ell_{i1,2}/c,~i=1,\dots,n$). 
The generalizations of 
Eqs.(\ref{mu1},\ref{lim},\ref{tanh1}),
assuming $M_i  < m < M_{i+1}$ for some $i \in \{1,\dots,n \}$ are:
\begin{eqnarray}
&&\mu = \mu (m,M_1, \dots, M_n), ~~~\textrm{such that,} \\
&& \mu = m + 
%\sum_{k=i+1}^{n} m_{0k} \cdot {\cal O} \left(\frac{m}{M_k} \right) +
\sum_{k=1}^{i} m_{0k} \cdot {\cal O} \left(\frac{M_k}{m} \right)
+  \sum_{k=i+1}^{n} m_{0k} \cdot {\cal O} \left(\frac{m}{M_k} \right)~, 
\nonumber \\
&&  \mbox{for}~\frac{m}{M_i} \gg 1,% \rightarrow \infty, 
~\frac{m}{M_{i+1}} \ll 1 %\rightarrow 0 
%\label{lim1b} \\
%
% && \mu = m + m_0 \cdot {\cal O} \left(\frac{M}{m} \right)~,~\mbox{as}~
% \frac{m}{M} \rightarrow \infty~, \label{lim2b} \\
%&&
%
;~\frac{m_{0k}}{M_k} \ll 1~,~k=1,..,n. 
\label{lim3b} \\
&& \mu (m,M_1,\dots,M_n) = \nonumber \\
&& \sum_{k=1}^n 
m_{0k} \left[ \tanh \left( \frac{m}{M_k} \right)^{a_k} \right] 
\sum_{j=1}^\infty d_{j_k} \left( \frac{M_\mathrm{k}}{m} \right)^{j_k}~,
\label{tanh2} 
%
%&& \mu = m + m_0 \left( \frac{m}{M_\mathrm{}} \right)^a 
%\left[ 1 - \frac{1}{3} \left( \frac{m}{M} \right)^2 + \cdots \right] \nonumber \\
%
%&& \times {\sum_{j=1}^{\infty} d_j \left( \frac{M_\mathrm{}}{m} \right)^j}, 
%\mbox{as}~~{\frac{m}{M} \rightarrow 0}
%\label{lim3c} \\
%
%&& 
%\mu = m + m_0 \left[  1 + 2 \sum_{l=0}^\infty (-)^l e^{- \frac{2lm}{~M} }
%\right] 
%\sum_{j=1}^{\infty } d_j \left( \frac{M}{m} \right)^j \nonumber \\
%
%&& \mbox{as}~\frac{m}{M} \rightarrow \infty~. \label{lim4c} 
\end{eqnarray}
with $a_k>1 \in \mathbb{R},~d_{j_k} =0, \forall j_k>a_k$.
This represents a series of mass scales starting from 
%above the 
%electroweak scale 
sub-Planckian energies
and stretching all the way to ultra high energy scales, 
with the Planck scale somewhere in between. Scales exceeding the 
$4$-dimensional Planck scale could for example represent  
quantum gravity scales in $5,6,\dots$ dimensions
(for this interpretation, one would have to use the Newton's constant 
in respective dimension to arrive at the $\ell_{i2}$ and $t_{i2}$ scales).
This shows our formalism may be of useful in theories involving 
extra dimensions, such as string theory or brane world models. 
Note that now one has $n$ new tiny mass scales 
$m_{0i}~(i=1,\dots,n)$, which are 
accessible only via gravitational interactions.

To conclude, we have shown in this paper that one or more invariant mass or length scales can be incorporated within the standard Lorentz transformations.
%\tcr{This guarantees that the standard dispersion relation between energy and momentum holds.}
Thus although physical energies still scale with Lorentz transformations, 
eigenvalues of geometrical operators are Lorentz invariant.
Furthermore this scale can, but need not be the Planck scale.
%; it can be any energy scale
%exceeding the electroweak scale.
%
The formalism can be applied to objects of any mass, small or big.
{Furthermore, this guarantees that the standard dispersion relation between energy and momentum holds,
and non-standard dispersion relations such as Eq.(\ref{dsrdisp})
or Eq.(\ref{disp3}) 
emerge as effective low-energy relations, without
any associated `soccer-ball' type problem.}
The scale shows up only when gravitational effects are called into play. 
This is analogous to e.g. the spin of an elementary particle such as an electron, which show up only in certain experiments (such as the Stern-Gerlach experiment) and not in others. Note that our proposal necessitates the existence of the mass scale $m_0$
(or the set $\{m_{0k}\}$), also with potential experimental or observational consequences.
The trade-off is a small but acceptable violation of the equivalence principle, which 
can also have observational signatures. 
Note that unlike in many previous studies, where the ratio of inertial
to gravitational mass was assumed to depend on the composition of the body, 
here we postulate the ratio to depend only on the magnitude of 
$m$ (or $\mu$), but {\it not} on its constituents. 
Furthermore, our proposal provides a theoretical basis for 
the relation between inertial and gravitational mass assumed in the literature,
e.g. Eq.(1) of \cite{will}, and shows that the latter holds for all bodies irrespective of size and composition. 
We claim that this is simpler and sheds light on the origin of inertia of a body. 

Future research in this direction should involve 
studying experimental implications of this proposal, both in classical 
(e.g. E\"otv\"os type experiments) and quantum 
(e.g. adapting our proposal to the experiment described in \cite{brukner},
using the Dirac equation, 
%\tcr
{studying collisions of elementary particles etc.}) 
systems, and estimating 
tighter bounds on $m_0$ and $M$. It is even possible that a new scale, much smaller than 
$M_P$, will be supported by experiments.
The origin of the mass $m_0$ (or $m_{0k}$ for multiple scales), for example 
in a Higgs mechanism in gravity
\cite{ssb1,ssb2} and
a reformulation of general relativity with $\mu\neq m$ is also in order. 
We hope to report on these issues elsewhere.

\section*{Acknowledgments}

This work was supported in part by the Natural Sciences and Engineering
Research Council of Canada. 
We thank the anonymous Referee for useful comments 
which have helped improve the paper. 

\section{References}

%\begin{thebibliography}{000} %for 3 digits
%\begin{thebibliography}{00}  %for 2 digits

\end{document}